\documentclass[letterpaper, 11pt]{article}
\usepackage[affil-it]{authblk} 
\usepackage{etoolbox}
\usepackage{lmodern}
\usepackage[letterpaper]{geometry}
\usepackage[in]{fullpage}
\usepackage[utf8]{inputenc}
\usepackage{amsmath}
\usepackage{amssymb}
\usepackage{cite}
\usepackage{tikz}
\usetikzlibrary{decorations.markings}
\usetikzlibrary{patterns}

\usepackage{soul}

\numberwithin{equation}{section}

\usepackage{footnote}
\usepackage{amsfonts}
\usepackage{graphicx}
\usepackage[colorlinks]{hyperref}
\hypersetup{linkcolor=blue,citecolor=blue,urlcolor=blue}

\usepackage{xcolor}
\usepackage{graphicx}
\usepackage{cancel}

\usepackage{subfiles}

\DeclareMathAlphabet{\mathpzc}{OT1}{pzc}{m}{it}
\DeclareMathAlphabet{\mathcalligra}{T1}{calligra}{m}{n}

\usepackage[framemethod=TikZ]{mdframed}
\mdfdefinestyle{MyFrame2}{%
	linecolor=blue!20,
	middlelinewidth=2pt,
	frametitlerule=true,
	frametitlebackgroundcolor=blue!20,
	frametitlerulecolor=blue!20,
	frametitlerulewidth=1pt,
	innertopmargin=\topskip}

\title{Brane motion in a compact space:\linebreak adiabatic perturbations of brane-bulk coupled fluids}

\makeatletter
\patchcmd{\@maketitle}{\LARGE \@title}{\fontsize{16}{19.2}\selectfont\@title}{}{}
\makeatother

\author[1]{Heliudson Bernardo\footnote{Email: \href{mailto:heliudson@hep.physics.mcgill.ca}{heliudson@hep.physics.mcgill.ca}}}

\author[1]{Fangyi Guo\footnote{Email: \href{mailto:fguo@physics.mcgill.ca}{fguo@physics.mcgill.ca}}}

\affil[1]{Department of Physics, McGill University, Montreal, QC, H3A 2T8, Canada}

\date{\vspace{-5ex}}

\begin{document}

\tikzset{middlearrow/.style={
        decoration={markings,
            mark= at position 0.5 with {\arrow{#1}} ,
        },
        postaction={decorate}
    }
}

\maketitle

\begin{abstract}
 
When a brane is moving in a compact space, bulk-probing signals originating at the brane can arrive back at the brane outside the lightcone of the emitting event. In this letter, we study how adiabatic perturbations in the brane fluid, coupled to a bulk fluid, propagate in the moving brane. In the non-dissipative regime, we find an effective sound speed for such perturbations, depending on the brane and bulk fluid energy densities, equations of state, and brane speed. In the tight-coupling approximation, the effective sound speed might be superluminal for brane and bulk fluids that satisfy the strong energy condition. This has immediate consequences for brane-world cosmology models.

\end{abstract}

\section{The relativity of motion in a compact space}

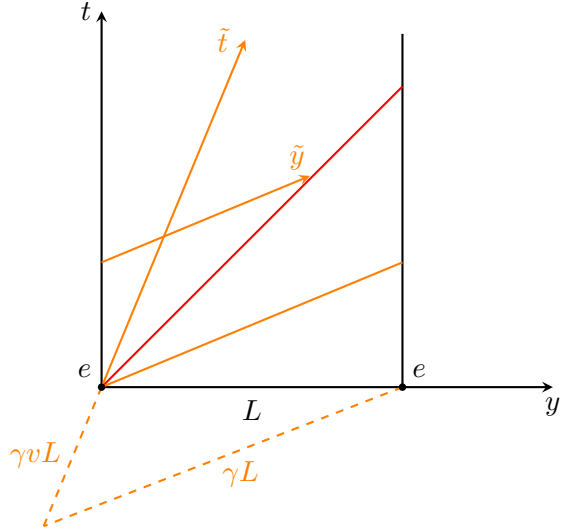
\begin{figure}[t]
    \centering
    \begin{tikzpicture}
        \coordinate[label = above left:$e$] (e0) at (0,0);
        \coordinate[label = above right:$e$] (e1) at (4,0);
        \coordinate[label=left:$t$] (t) at (0,5);
        \coordinate[label=below:$y$] (y) at (6,0);
        \draw[thick, black, -stealth] (0,0) -- (0,5);
        \node at (t) {};
        \draw[thick, black, -stealth] (0,0) -- (6,0);
        \node at (y) {};
        \node at (2,-0.3) (L) {\(L\)};
        \draw[thick, black] (4,0) -- (4,4.7);
        \draw[thick, red] (0,0) -- (4,4);
        \draw[thick, orange, -stealth] (0,0) -- (1.91,4.62);
        \node[orange] at (1.61,4.62) (tTil) {\(\Tilde{t}\)};
        \draw[thick, orange] (0,0) -- (4,1.66);
        \draw[thick, orange, -stealth] (0,1.66) -- (2.77,2.81);
        \node[orange] at (2.60,3.05) (yTil) {\(\Tilde{y}\)};
        \draw[dashed, thick, orange] (0,0) -- (-0.77,-1.85);
        \node[orange] at (-0.88,-0.9) (GammaVL) {\(\gamma vL\)};
        \draw[dashed, thick, orange] (4,0) -- (-0.77,-1.85);
        \node[orange] at (1.85,-1.15) (GammaL) {\(\gamma L\)};
        \node at (e0) [circle,fill,inner sep=1pt]{};
        \node at (e1) [circle,fill,inner sep=1pt]{};
    \end{tikzpicture}
    \caption{Spacetime diagram of the motion of the frame $K_v$ in $K_0$ coordinates, with $v>0$. The lines $y= 0$ and $y= L$ are identified. The event $e$ at the origin in both frames can also be described by the coordinates $(\tilde{t}, \tilde{y}) = (-\gamma v L, \gamma L)$. This can be seen from the dashed red lines extended from the fundamental region $0\leq y<L$. In the frame $K_v$, the total size of the compact direction is $\gamma L$.}
    \label{fig1}
\end{figure}

Consider a five-dimensional Minkowski spacetime with one spatial direction compactified to a circle, $\mathcal{M}_5 = \mathcal{M}_4 \times S^1$. Suppose that for an inertial frame $K_0$, coordinates $x^\mu = (t, x^i, y)$ are assigned to events, where $y$ is the compact direction, and the coordinates of an event with $y= y_0$ and $y=y_0 + L$ should be identified, i.e. $(t, x^i,y) \sim (t, x^i,y + L)$, where $L$ is the total length of the compact direction as measured in such a frame. For any other inertial frame $K_v$ that moves with respect to $K_0$ with speed $v$ in the $y$-direction, the identification will fail to be purely spacelike: the coordinates $\Tilde{x}^\mu = (\Tilde{t}, \Tilde{x}^i, \Tilde{y})$ of an event in $K_v$ will be identified as $\tilde{x}^\mu \sim (\Tilde{t} - \gamma vL, \Tilde{x}^i,\Tilde{y}+ \gamma L )$, as follows from a Lorentz boost in the coordinates of the identified event, see also the spacetime diagram of Figure \ref{fig1}. 

\begin{figure}[t]
    \centering
    \begin{tikzpicture}
        \coordinate[label=left:$\Tilde{t}$] (tTil) at (0,5);
        \coordinate[label=below:$\Tilde{y}$] (yTil) at (6,0);
        \coordinate[label=above left:$e$] (e0) at (0,0);
        \coordinate[label=below left:$e$] (e1) at (3,-1);
        \coordinate[label=right:$f_2$] (f2) at (3,3);
        \coordinate[label=above right:$f_1$] (f1) at (0,2);
        \coordinate[label=above right:$f_2$] (f22) at (0,4);
        \draw[thick, black, -stealth] (0,0) -- (tTil);
        \node at (tTil) {};
        \draw[thick, black, -stealth] (0,0) -- (yTil);
        \node at (yTil) {};
        \draw[thick, black] (3,-2) -- (3,4.7);
        \draw[thick, orange, middlearrow={stealth}] (e0) -- (f2);
        \draw[thick, orange, middlearrow={stealth}] (e1) -- (f1);
        \draw[|<->|] (0,-0.2) -- (3,-0.2);
        \node at (1.5,-0.5) (gammaL) {$\gamma L$};
        \draw[|<->|] (3.2,0) -- (3.2,-1);
        \node at (3.7,-0.5) (gammaVL) {$\gamma vL$};
        \draw[|<->|] (-0.2,2) -- (-0.2,4);
        \node at (-0.8,3) (2gammaVL) {$2\gamma vL$};
        \node at (e0) [circle,fill,inner sep=1pt]{};
        \node at (e1) [circle,fill,inner sep=1pt]{};
        \node at (f2) [circle,fill,inner sep=1pt]{};
        \node at (f1) [circle,fill,inner sep=1pt]{};
        \node at (f22) [circle,fill,inner sep=1pt]{};
    \end{tikzpicture}
    \caption{Global experiment to probe $K_v$'s inertial motion. The line $\Tilde{y} =0$ is identified with the time-translated $\Tilde{y} = \gamma L$ line. Hence a lightray emitted from event $e$ in the negative-$\Tilde{y}$ direction arrives back to $\Tilde{y}=0$ at $f_1$ while the lightray emitted in the frame's motion direction arrives back to $\Tilde{y}=0$ at $f_2$. Since the identified coordinates are time-translated by $\gamma vL$, the difference between arrival times is  $2\gamma vL$. Note that signals with propagating speeds other than $c$ also undergo time advancement or delay when crossing the compact direction.}
    \label{fig2}
\end{figure}
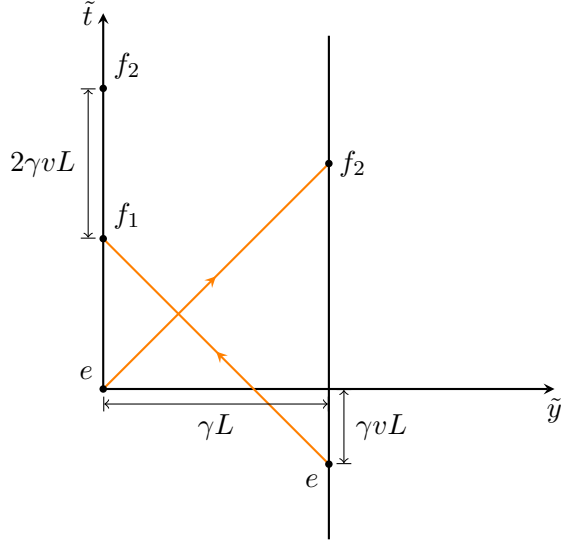

The time contribution to the identification results from the relativity of simultaneity between the frames because it is only in $K_0$ that the identified events are simultaneous. There is only one frame in which the total size of the compact direction is $L$, while the measurable size in any other frame moving relative to it will be $\gamma L$ \cite{Peters:1983as}. Moreover, in the frame $K_v$, if light signals are sent towards and backward the direction of motion, there is a mismatch $\Delta t = 2\gamma v L$ between their arrival times back at the emission point, see Figure \ref{fig2}. Thus, there is a preferred frame in $\mathcal{M}_4\times S^1$, and inertial observers can perform experiments to detect its state of motion relative to it, provided these probe the global structure of the spacetime. This is a manifestation of the global breaking of Lorentz symmetry by compactification. Other consequences of the relativity of identifications between the frames were explored in \cite{Peters:1983as}, such as the impossibility of global synchronization of clocks. For other works about relativity on compact spaces, see \cite{Barrow:2001rj,Uzan_2002,Luminet:2009xw} and references therein.

If an observer in $K_v$ wants to send a prompt signal to an event with spatial separation $r$ along $\tilde{y} =\text{constant}$, she could do it along $\tilde{y} = \text{constant}$ or along a direction with an angle $\theta = \arctan\left(\gamma L/r\right)$ relative to $\tilde{y} = \text{constant}$. If $r/\gamma L\ll 1$, the former option is always more prompt. However, for a sufficiently large $r/\gamma L$, the signal propagating with angle $\theta$ can arrive \emph{earlier} than the signal that does not probe the compact direction. In fact, in $K_v$, for a signal propagating at speed $u$, the time it takes to get at distance $r$ along the $\theta$-direction is
\begin{equation}\label{tadv}
    \Delta t = \frac{\gamma L}{u \sin{\theta}} - \gamma v L,
\end{equation}
where the second term comes from the time advancement after the signal crosses the compact direction (see Figure \ref{fig2}). This can be smaller than the time $r/u$ that the signal would take to cover $r$ along $\tilde{y} = \text{constant}$ provided
\begin{equation}
    \frac{r}{\gamma L} > \frac{1}{2vu}\left(1-v^2 u^2\right).
\end{equation}
Note that when $v = 0$, signals that probe the compactification are always less prompt than signals that do not. 

Recently, in \cite{Greene_2022}, a 3-brane moving in a compact direction was studied, and the compacti-\linebreak fication-induced Lorentz symmetry breaking was exploited to show that brane-originated, bulk-propagating signals can beat brane propagating signals in the promptness of information transmission. This was explored in detail in \cite{Greene_2022} (see also \cite{Greene:2011fm}), after studying the retarded Green's function of a scalar field on $\mathcal{M}_4\times S^1$ in different frames. Due essentially to the time advancement in \eqref{tadv}, the motion of the brane modifies brane causality such that signals probing the bulk can get back to the brane outside the brane's light cone, which is an effective superluminal propagation between events on the brane. Note, however, that there is no causality violation since the time advancement that a bulk-propagating signal can acquire is always smaller than the time light takes to cross the bulk \cite{Peters:1983as, Greene:2022uyf}.

\begin{figure}[t]
    \centering
    \begin{tikzpicture}
        \coordinate[label=left:$\Tilde{t}$] (tTil) at (0,2);
        \coordinate[label=below:$\Tilde{y}$] (yTil) at (2,0);
        \coordinate[label=below:$\Tilde{r}$] (rTil) at (-1.73,-1);
        \coordinate[label=below left:$e$] (e0) at (3.7,1.5);
        \coordinate (p1) at (3.4,3.0);
        \coordinate (p2) at (4.4,3.6);
        \coordinate (p3) at (2,3.2);
        \coordinate (p4) at (2.8,3.66);
        \coordinate (p03) at (3,2.2);
        \coordinate (p13) at (3,3);
        \coordinate (p24) at (3,3.72);
        \coordinate[label={[label distance=-3pt]300:$e$}] (e1) at (2.48,2.35);
        \coordinate (p5) at (2.1,3.26);
        \coordinate (p6) at (2.7,3.60);
        \draw[thick, black, -stealth] (0,0) -- (tTil);
        \node at (tTil) {};
        \draw[thick, black, -stealth] (0,0) -- (yTil);
        \node at (yTil) {};
        \draw[thick, black, -stealth] (0,0) -- (rTil);
        \node at (rTil) {};
        \draw[thick] (3,0.7) -- (5.60,2.2);
        \draw[thick] (3,0.7) -- (3,4.7);
        \draw[thick] (3,4.7) -- (5.60,6.2);
        \draw[thick] (5.60,2.2) -- (5.60,6.2);
        \node at (e0) {};
        \draw[thick, orange] (e0) -- (p1);
        \draw[thick, orange] (e0) -- (p2);
        \draw[thick, dashed, orange] (e0) -- (p03);
        \draw[thick, orange, dashed] (p1) .. controls (3.1,2.97) .. (p13);

        \draw[thick] (1.21,0.7) -- (3,1.73);
        \draw[thick] (1.21,0.7) -- (1.21,4.7);
        \draw[thick] (1.21,4.7) -- (3.81,6.2);
        \draw[thick] (3.81,6.2) -- (3.81,5.17);
        \draw[thick, orange] (p03) -- (p3);
        \draw[thick, orange] (p13) .. controls (2.8,2.98) and (2.2,3.1) .. (p3);
        \draw[thick, orange] (p3) .. controls (2.3,3) and (2.7,2.7) .. (p4);
        \draw[thick, orange] (p4) -- (p24);
        \draw[thick, orange, dashed] (p2) .. controls (3.9,3.75) and (3.4,3.8) .. (p24);

        \node at (e1) {};
        \draw[thick, red] (e1) -- (p5);
        \draw[thick, red] (e1) -- (p6);
        \draw[thick, red] (p5) .. controls (2.3,3.15) and (2.7,3.2) .. (p6);
    \end{tikzpicture}
    \caption{Future lightcone emanating from an event $e$. The radial coordinate of the directions transverse to the brane's motion $r = \sqrt{x^i x_i}$ is also shown. After some time, the maximal distance $e$ can influence is set by the lightcone emanating from the $\Tilde{y} = \gamma L$ plane. This can be seen from the portion of the red light cone emanating from the $\Tilde{y}=0$ which is contained in the yellow lightcone. Assymptotically, the future light-``cone'' of $e$ is set by the hyperbola resulting from the intersection of the yellow cone with the $\Tilde{y} = 0$ plane.}
    \label{fig3}
\end{figure}
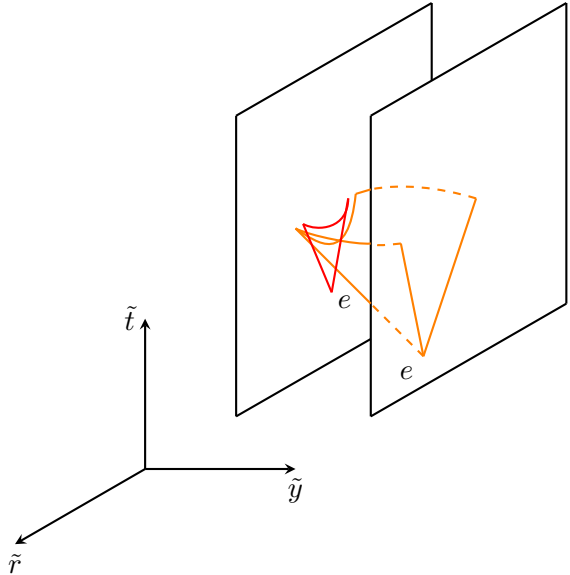

The spacetime diagram in Figure \ref{fig3}, shows the lightcone of an event on the brane and how it spreads through the compact direction, in coordinates comoving with the brane. Initially, the intersection of the brane with the lightcone emanating from a brane event $e$ is described by $r(\tilde{t}) = \tilde{t}$, where $r$ is the radial coordinate of the directions parallel to the brane. However, after $\Delta t = \gamma L - \gamma v L$, the ligthcone intersects the brane at the hyperbola
\begin{equation}
    r^2(\tilde{t}) = (\tilde{t}+\gamma v L)^2 - \gamma^2 L^2,
\end{equation}
and for $\tilde{t}>L/2\gamma v$, the naive ``brane lightcone'' is contained in the genuine lightcone that probes the bulk. Moreover, asymptotically in time, the difference between these approaches a constant, $r(\tilde{t})-r_\parallel(\tilde{t}) \to \gamma v L$. 

An interesting application of the effective superluminal propagation is the possibility of having a solution to the cosmological horizon problem in brane-world models. This was already suggested in \cite{Greene_2022}, with a brief discussion on how the brane particle horizon would change if the brane is moving. The horizon problem is a causality problem due to the finiteness of the distance photons could have traveled from the Big Bang until recombination. In standard Big-Bang cosmology, such a distance is much smaller than the particle horizon at recombination \cite{Brandenberger:1999sw, Baumann:2022mni}. So, the isotropy of the CMB thermal spectrum cannot be explained by a causal mechanism within the $\Lambda$CDM model. However, if one supposes that our Universe is a 3-brane moving in a compact direction, then regions outside the brane's particle horizon could have been in causal contact, provided there is a coupling between the brane's primordial plasma and bulk propagating fields. One could imagine such a coupling being relevant only before Big Bang Nucleosynthesis, so cosmological observables are mildly affected by the bulk coupling.

In this work, we study how a brane-bulk coupling modifies the propagation of perturbations in a 3-brane that is moving in a compact direction. We consider brane and bulk-perfect fluids coupled by an electric-like coupling in the tight-coupling approximation, such that dissipation and non-adiabatic perturbations are subleading effects. The system has similarities with the photon-baryon fluid of the primordial plasma deep in the radiation-dominated era. We are interested in the consequences of the minimal assumption that there is a coupling between brane and bulk fluids, as required to thermalize otherwise causally-disconnected brane regions. In the next section, we find the equation of motion for adiabatic perturbations in the moving brane, including the bulk coupling. In section \ref{sec3}, we comment on the physics behind our assumptions on the brane-bulk interactions. Section \ref{sec4} is a discussion about how cosmological expansion would modify our results. We discuss its implications in the conclusion section.

\section{Interacting brane and bulk-fluids}
\label{sec2}

We consider a 3-brane moving with speed $v$ in an extra transverse direction which is compactified to a circle. In this section, we shall assume a flat spacetime. In the frame comoving with the brane, the metric is given by
\begin{equation}
    ds^2 =-dt^2 + dx^i dx_i + dy^2,
\end{equation}
 where $y$ is the compact direction. The presence of the compact direction does not affect local energy-momentum conservation,
 \begin{equation}
     \nabla_M T^{MN} =0,
 \end{equation}
where $T_{MN}$ is the five-dimensional energy-momentum tensor. It receives contributions from the bulk and brane fluids:
\begin{equation}
    T^{MN} = T^{MN}_{\text{brane}} + T^{MN}_{\text{bulk}}.
\end{equation}
In order for brane signals to exploit the bulk direction and arrive at the brane outside the initial point's lightcone, we will assume that the bulk and brane fluids are coupled,
\begin{equation}\label{braneandbulkconservation}
    \nabla_M T^{MN}_{\text{brane}} = J^N, \quad \nabla_M T^{MN}_{\text{bulk}} = -J^N.
\end{equation}
The coupling vector $J^N$ is localized on the brane and proportional to the bulk and brane fields' interaction rate $\Gamma$. We consider elastic interaction only, such that each fluid's particle number is individually conserved. 

Moreover, we shall assume the bulk fluid to be a perfect fluid,
\begin{equation}
    T^{MN}_{\text{bulk}} = \rho_b U^M_b U^N_b + p_b (U^M_b U^N_b + g^{MN}),
\end{equation}
while the brane fluid is also a perfect fluid,
\begin{equation}
    T^{MN}_{\text{brane}} = \rho U^M U^N + p(U^M U^N + g^{MN}) + (p_4 - p) X^M X^N,
\end{equation}
 but with $\rho$, $p$ and $U^M$ localized on the brane: they all have factors of $\delta(y)$ (assuming the brane is at $y=0$), and the brane-fluid velocity field has vanishing $y$ component, $U^4 = 0$. Moreover, $X^M$ is a spacelike unit vector transverse to $U^M$, and there is no pressure in the extra direction, $p_4=0$. In the frame comoving with the brane, we have $X^M = (0,0,0,0,1)$ and $U_b = \gamma(1, u_b^i, -v)$, where $v$ is the brane speed in the frame comoving with the bulk, i.e., the special frame where the coordinate identification is purely spatial. 
 
In the following, we write the continuity and Euler equations for the brane and bulk fluids. We find a background solutions after assuming that the energy-momentum transfer flux $J^N$ is of first order in perturbations. We comment more about $J^N$ after finding the equations for linearized perturbations. From \eqref{braneandbulkconservation}, the brane continuity and Euler equations are
\begin{subequations}\label{branefluid}
\begin{align}
        & U^M \nabla_M \rho + (\rho + p) \nabla_M U^M = - J_N U^N,\\
        &g^{QM}\nabla_M p + U^Q U^M\nabla_M p + (\rho + p)U^M \nabla_M U^Q + X^Q \left[X^M\nabla_M (p_4-p) + (p_4 - p)\nabla_M X^M\right] \nonumber \\
        &+(p_4 -p)\left[X^M \nabla_M X^Q + U^Q U^N X^M \nabla_M X_N\right] = J^Q + U^Q J_N U^N.
\end{align}
\end{subequations}
In the frame comoving with the fluid parallel to the brane, we have $U^M = (1,0,0,0,0)$. Using this in the brane continuity and Euler equations yields
\begin{subequations}\label{branegeneralmectric}
\begin{align}
    \partial_0 \rho &= -J_0,\\
    \partial^i p &= J^i + J_0 u^i, \\
    \partial^4 p + \partial_4 (p_4 - p) &= J^4.
\end{align}
\end{subequations}
If we assume that $J^N$ vanishes at the background level, we find the solution
\begin{equation}\label{branebacksol}
    \rho = \rho_0 \delta(y), \quad p = p_0\delta(y), \quad p_4 = 0,
\end{equation}
where $\rho_0$ and $p_0$ are constants. 

The bulk continuity and Euler equations are
\begin{subequations}\label{bulkfluid}
    \begin{align}
        U^M_b \nabla_M \rho_b + (\rho_b + p_b) \nabla_M U^M_b &= J_N U^N_b,\\
        g^{QM}\nabla_M p_b + U^Q_b U^N_b\nabla_N p_b + (\rho_b + p_b)U^M_b \nabla_M U^Q_b &= -J^Q - U^Q_b J_N U^N_b.
    \end{align}
\end{subequations}
We have $U^M = \gamma(1, u^i_b, -v)$ in the brane comoving frame, such that these equations reduce to 
\begin{subequations}\label{bulkfluidopen}
    \begin{align}
        \gamma \partial_0 \rho_b + \gamma u^i_b \partial_i \rho_b - \gamma v \partial_4 \rho_b +(\rho_b+ p_b)\left[\partial_0 \gamma + \partial_i (\gamma u^i_b) - \partial_4(\gamma v)\right] &= J_NU^N_b, \\
        \partial^i p_b + \gamma^2 u^i_b (\partial_0 + u^i_b \partial_i -v \partial_4)p_b 
        +(\rho_b +p_b)\left[\gamma (\partial_0 + u^i_b \partial_i -v \partial_4)(\gamma u^i_b) \right] &= -J^i - \gamma u^i_b J_NU^N_b, \\
        \partial^4 p_b - \gamma^2 v(\partial_0 + u^i_b \partial_i -v \partial_4)p_b  +(\rho_b +p_b)\left[-\gamma(\partial_0 + u^i_b \partial_i -v \partial_4)(\gamma v)\right] &= -J^4 + \gamma v J_N U^N_b,
    \end{align}
\end{subequations}
with $J_N U^N_b = \gamma(J_0 + J_i u^i_b - J_4 v)$. For vanishing $J^N$, we find the solution
\begin{equation}\label{bulkbacksol}
    \rho_b = \rho_{b,0} = \text{const.}, \quad p_b = p_{b,0} = \text{const.}, \quad u^i_b = 0, 
\end{equation}
and constant $v$.

The equations for brane perturbations can be obtained from \eqref{branefluid} after setting $\rho = \bar{\rho} + \delta \rho$, $p = \bar{p} + \delta p$, $p_4 = \Bar{p}_4 + \delta p_4$, and $U^M = (1, u^i, 0)$, with $|u^i| \ll 1$. Note that we impose these perturbations to be restricted to the brane, so that $U^4$ is still zero and $\delta \rho$ and $\delta p$ are also localized on the brane.  Then, to linear order in $\delta \rho$, $\delta p$, and $u^i$, these brane perturbations satisfy,
\begin{subequations}
\label{branepertafterbackgr}
    \begin{align}
        \partial_0 \delta \rho + (\Bar{\rho} +\Bar{p}) \partial_i u^i &= -J_0,\\
        (\Bar{\rho}+ \Bar{p})\partial_0 u^i + \partial^i \delta p &= J^i,\\
        \partial_y \delta p_4 &= J^4. \label{branepertafterbackgr3}
    \end{align}
\end{subequations}
where the background quantities $\Bar{\rho}$ and $\Bar{p}$ are evaluated in the solution \eqref{branebacksol}.

For bulk perturbations, we can already use \eqref{bulkfluidopen}, but with $v = \Bar{v}+ \delta v$, $\rho_b = \Bar{\rho}_b + \delta \rho_b$, $p_b = \Bar{p}_b +\delta p_b$, and treating $u^i_b$
as a perturbation. So, $\gamma  = \Bar{\gamma} + \delta \gamma$ with $\Bar{\gamma} = (1-v^2)^{-1/2}$ and $\delta \gamma = \Bar{v}\delta v\Bar{\gamma}^3$ to linear order. Setting $\Bar{\rho}_b$, $\Bar{p}_b$, and $\Bar{v}$ to the background solution \eqref{bulkbacksol}, gives
\begin{subequations}\label{bulkpertafterbackgr}
    \begin{align}
        \Bar{\gamma}(\partial_0 - \Bar{v} \partial_y)\delta \rho_b + (\Bar{\rho}_b+ \Bar{p}_b)\left[\partial_0 \delta \gamma + \Bar{\gamma}\partial_i (u^i_b)- \Bar{v} \partial_4 \delta \gamma - \Bar{\gamma}\partial_4 \delta v\right] &= J_0\Bar{\gamma} - J_4 \Bar{\gamma}\Bar{v},\\
        \Bar{\gamma}^2(\Bar{\rho}_b+\Bar{p}_b)\left(\partial_0- \Bar{v}\partial_4\right)u^i_b + \partial^i \delta p_b  &= -J^i, \\
        -\Bar{\gamma}(\Bar{\rho}_b + \Bar{p}_b)\left(\partial_0 -\Bar{v}\partial_4\right)(\delta \gamma \Bar{v}+ \Bar{\gamma}\delta v) + \partial^4 \delta p_b -\Bar{\gamma}^2\Bar{v}(\partial_0 - \Bar{v}\partial_4)\delta p_b &= - J^4 + (J_0 - \Bar{v}J_4)\Bar{\gamma}^2 \Bar{v}.
    \end{align}
\end{subequations}
Note that since $J^N$ is a distribution, so will be the bulk fluid perturbation variables. This can be seen already from \eqref{braneandbulkconservation}: the second equation implies a discontinuity in the bulk energy-momentum tensor across $y=0$. However, since we are interested in the bulk fluid variables evaluated on the brane, we shall not explicitly write $\delta (y)$ factors. 

In $\Lambda$CDM, during radiation domination, the radiation and matter cosmic fluids are coupled mainly due to Thomson scattering. This interaction enters the Boltzmann equation describing the phase-space distribution of the fluid's particles. In the hydrodynamic limit, the result is an energy-momentum transfer flux that vanishes at the background level and is proportional to the difference between the fluid's velocity perturbations \cite{Hu:1995en, Weinberg:2008zzc}. This can also be seen from a covariant formalism, as shown in \cite{Uzan:1998mc}. We shall assume the same behaviour for $J^N$. Hence it should be proportional to the perturbations in the brane and bulk velocities:
\begin{equation}
    J^N = \Theta (\delta U^M - \delta U^M_b),
\end{equation}
 where $\delta U^M = (0, u^i, 0)$, $\delta U^M_b = (\delta \gamma, \Bar{\gamma} u^i_b, - \Bar{\gamma}\delta v)$, and $\Theta$ depends on the cross-section $\sigma$ of the interaction between the brane and bulk particles. For simplicity, we impose the form of $J^N$ above and postpone discussion about its microphysics description to elsewhere. Hence, we have
 \begin{equation}\label{energymomentumtransfer}
     J^0 = -\Theta \delta \gamma, \quad J^i= \Theta (u^i - \Bar{\gamma}u^i_b), \quad J^4 = - \Theta \Bar{\gamma} \delta v.
 \end{equation}

We now wish to find a closed equation for the brane energy density perturbation $\delta \rho$. To accomplish this, we need to combine equations \eqref{branepertafterbackgr} and \eqref{bulkpertafterbackgr}. Before we do so, we impose another simplyfying assumption: we set the perturbation in the pressure $p_4$ to zero, $\delta p_4 = 0$. This is justified by the fact that we are interested in perturbations which are confined to the brane. If we do so, equation \eqref{branepertafterbackgr3} gives $J^4 = 0$, which from \eqref{energymomentumtransfer} implies $\delta v =0$ and $J^0 = 0$. So, to this order in approximation, the brane-bulk scattering process does not allow energy exchange, only momentum. Therefore, to leading order, the continuity equations for the brane and bulk perturbations are not modified by the scattering, while the Euler equations do get contributions from the interaction. In the next section, we explain the relation between $J^0=0=J^4$, dissipation, and adiabatic perturbations.

With $\delta v= 0$, and assuming a linear barotropic relation between the energy densities and pressures $p = w \rho$, $p_b = w_b \rho_b$, \eqref{branepertafterbackgr} imply
\begin{subequations}
    \begin{align}
        \partial_0 \delta + (1+ w)\partial_i u^i &= 0, \\
        (1+w)\partial_0 u^i + c^2_s \partial^i \delta &= \frac{\Theta}{\Bar{\rho}}(u^i - \Bar{\gamma} u^i_b),
    \end{align}
\end{subequations}
where $\delta = \delta\rho/\Bar{\rho}$ and $\delta p = c^2_s \delta \rho$, with $c_s^2 = w$ the usual sound speed for brane perturbation if it were no interaction with the bulk. Meanwhile, equations \eqref{bulkpertafterbackgr} can be written as
\begin{subequations}
    \begin{align}
        (\partial_0 - \Bar{v} \partial_4)\delta_b + (1+w_b) \partial_i u^i_b &= 0,\\
        \Bar{\gamma}^2(1+w_b)(\partial_0 - \Bar{v}\partial_4)u^i_b + c_b^2 \partial^i \delta_b &= -\frac{\Theta}{\Bar{\rho}_b}(u^i - \Bar{\gamma}u^i_b),\\
        c^2_b \partial^4 \delta_b - \Bar{\gamma}\Bar{v}(\partial_0 -\Bar{v}\partial_4)\delta_b &= 0,
    \end{align}
\end{subequations}
where $\delta p_b = c^2_b \delta \rho_b$.

From the Euler equation for the bulk fluid, we find a relation between $u^i$ and $u^i_b$. In the tight-coupling approximation, where $u^i \approx u^i_b$ to first order, we find
\begin{equation}
    u^i- \Bar{\gamma}u^i_b \approx - \frac{\Bar{\rho}_b}{\Theta}\left[ \Bar{\gamma}^2(1+w_b)(\partial_0 - \Bar{v}\partial_4)u^i + c_b^2 \partial^i \delta_b \right].
\end{equation}
Inserting this into the Euler equation for the brane fluid gives
\begin{equation}
    (1+w)\partial_0 u^i + c_s^2 \partial^i \delta  = -\frac{\Bar{\rho_b}}{\Bar{\rho}}\left[\Bar{\gamma}^2(1+w_b)(\partial_0 - \Bar{v}\partial_4)u^i + c_b^2 \partial^i \delta_b\right].
\end{equation}
As explained in the next section, for adiabatic perturbations, we have $\delta_b = \alpha \delta$ with the constant $\alpha$ given by \eqref{alphaadiabaticpert}. Using this relation in the previous equation, we find
\begin{equation}\label{modifiedBraneEuler}
    \left[(1+w)+ \frac{\Bar{\rho}_b}{\Bar{\rho}} \Bar{\gamma}^2 (1+w_b)\right]\partial_0 u^i - \frac{\Bar{\rho}_b}{\Bar{\rho}}\Bar{\gamma}^2 \Bar{v}(1+w_b)\partial_4 u^i + c_s^2\left(1 + \alpha \frac{\Bar{\rho}_b c_b^2}{\Bar{\rho}c_s^2}\right)\partial^i \delta = 0.
\end{equation}
Taking $\partial_0$ of the brane's continuity equation gives
\begin{equation}
    \partial_0^2 \delta + (1+ w) \partial_0 \partial_i u^i = 0, 
\end{equation}
while the divergence of \eqref{modifiedBraneEuler} yields
\begin{equation}
    \partial_i \partial_0 u^i = \frac{\Bar{\rho}_b}{\Bar{\rho}}\frac{\Bar{\gamma}^2 \Bar{v}(1+w_b)}{1+w + \frac{\Bar{\rho}_b}{\Bar{\rho}}\Bar{\gamma}^2 (1+w_b)}\partial_i \partial_4 u^i - c_s^2 \frac{1+ \alpha \frac{\Bar{\rho}_bc_b^2}{\Bar{\rho}c_s^2}}{1+w + \frac{\Bar{\rho}_b}{\Bar{\rho}}\Bar{\gamma}^2 (1+w_b)}\partial_i \partial^i \delta.
\end{equation}
Combining the last two equations, we get
\begin{equation}
    \partial_0^2 \delta + \frac{\Bar{\rho}_b}{\Bar{\rho}} \frac{\Bar{\gamma}^2 \Bar{v}(1+w_b)}{1+ \Bar{\gamma}\frac{\Bar{\rho}_b(1+w_b)}{\Bar{\rho}(1+w)}} \partial_i \partial_4 u^i - c_s^2 \frac{1+ \alpha\frac{\Bar{\rho}_b c_b^2}{\Bar{\rho}c_s^2}}{1+ \Bar{\gamma} \frac{\Bar{\rho}_b(1+w_b)}{\Bar{\rho}(1+w)}} \partial_i \partial^i \delta =0.
\end{equation}
But since the perturbations are restricted to the brane, $\partial_4 u^i = 0$, such that we get the final equation for perturbations on the brane:
\begin{equation}\label{pertonbrane}
    \partial_0^2 \delta - c_s^2 \frac{1+ \alpha\frac{\Bar{\rho}_b c_b^2}{\Bar{\rho}c_s^2}}{1+ \Bar{\gamma} \frac{\Bar{\rho}_b(1+w_b)}{\Bar{\rho}(1+w)}} \partial_i \partial^i \delta =0.
\end{equation}
Thus, we find that, in the tight-coupling approximation, the sound speed for perturbations on the brane is
\begin{equation}\label{soundspeed1}
    c_{\text{eff}} = c_s \sqrt{\frac{1+ \alpha\frac{\Bar{\rho}_b c_b^2}{\Bar{\rho}c_s^2}}{1+ \Bar{\gamma} \frac{\Bar{\rho}_b(1+w_b)}{\Bar{\rho}(1+w)}}}.
\end{equation}
We shall discuss this result after commenting on the physical meaning of the assumptions on the interaction that we made in this section.

\section{Adiabatic perturbations and sound speed of the interacting \linebreak brane-bulk fluid system}
\label{sec3}

In this section, we show that $J^0= 0= J^4$ implies that $\delta$ and $\delta_b$ correspond to adiabatic perturbations. To better state the argument, consider a single relativistic fluid defined in an arbitrary spacetime. Local conservation of energy and momentum rules the fluid's dynamics,
\begin{equation}
    \nabla_M T^{MN} = 0,
\end{equation}
where $T^{MN}$ is the energy-momentum tensor of the fluid. For a general fluid, the interactions of the fluid particles might be such that there are dissipative effects making the fluid's entropy increase. For instance, when the particle collisions are not elastic, or the fluid has multiple components that exchange energy and momentum. Let's first assume that this is not the case and that the fluid is a perfect fluid with conserved particle number,
\begin{equation}\label{pnumberconservation}
    \nabla_M N^M = 0,
\end{equation}
where 
\begin{equation}
    T^{MN} = \rho U^M U^N + p(U^M U^N + g^{MN}), \quad N^M = n N^M,
\end{equation}
where $\rho$, $p$, and $n$ are the energy density, pressure, and number density in the frame comoving with the fluid, respectively. 

Given these assumptions and the first and second laws of thermodynamics on a volume $V$ on the fluid's frame, 
\begin{equation}
    T d\sigma = d \left(\frac{\rho}{n}\right) + p d\left(\frac{1}{n}\right),
\end{equation}
one can show from \eqref{pnumberconservation} and the continuity equation, $U_N \nabla_M T^{MN} =0$, that the entropy per particle number $\sigma = S/N$ (as measured in the fluid's rest frame) is constant along the fluid flow \cite{Weinberg:1972kfs}:
\begin{equation}
    U^M \nabla_M \sigma =0.
\end{equation}
Equivalently, in terms of the entropy density $s = S/V = n\sigma$, we have
\begin{equation}
    \nabla_M \left(sU^M\right)=0,
\end{equation}
such that, for a perfect fluid with conserved particle number density, entropy is conserved.

A general perturbation in the macrostate of the fluid includes a fluctuation in $\sigma$. Perturbations are adiabatic if $\delta \sigma =0$. For such a class of perturbations, there is a relation between the energy and number density fluctuations:
\begin{equation}
    0=T \delta \sigma = \frac{1}{n}\left(\delta \rho - \frac{\rho+p}{n}\delta n\right) \implies \delta \rho = \frac{\rho + p}{n}\delta n.
\end{equation}

Now, let's suppose the fluid is made of two interacting components but in such a way that the continuity and number density conservation equations are individually satisfied for both components. In this case, 
\begin{equation}
    T \delta \sigma_1 = \frac{1}{n_1}\left(
    \delta \rho_1 -\frac{\rho_1 + p_1}{n_1}\delta n_1\right), \quad T \delta \sigma_2 = \frac{1}{n_2}\left(
    \delta \rho_2 -\frac{\rho_2 + p_2}{n_2}\delta n_2\right),
\end{equation}
such that if the perturbations in each component are adiabatic,
\begin{equation}\label{adiabaticpertcomponents}
    \delta \rho_1 = \frac{\rho_1 + p_1}{n_1}\delta n_1, \quad \delta \rho_2 = \frac{\rho_2 + p_2}{n_2}\delta n_2.
\end{equation}
However, if we consider the system as a whole, with $\rho = \rho_1+ \rho_2$, $p = p_1+p_2$, and $n = n_1 + n_2$, the condition for the perturbations in the total fluid to be adiabatic is 
\begin{equation}
    \delta \rho_1 +\delta \rho_2 = \frac{\rho_1 + \rho_2 + p_1 +p_2}{n_1 + n_2}(\delta n_1 + \delta n_2).
\end{equation}
Using the expressions for $\delta \rho_1$ and $\delta \rho_2$ in \eqref{adiabaticpertcomponents} and making some algebraic manipulations, we find
\begin{equation}
    \frac{\delta n_1}{n_1} = \frac{\delta n_2}{n_2}.
\end{equation}
As a consistency check, the same condition can be found after imposing $\delta \sigma = 0$, because
\begin{equation}
    \delta \sigma = \delta \left(\frac{S_1 + S_2}{N_1+N_2}\right) = \frac{n_1 n_2}{(n_1+n_2)^2}\left(\sigma_1- \sigma_2\right)\left(\frac{\delta n_1}{n_1}- \frac{\delta n_2}{n_2}\right) + \frac{n_1 \delta \sigma_1 + n_2 \delta \sigma_2}{n_1+n_2}
\end{equation}
Thus, from \eqref{adiabaticpertcomponents}, we find a relation between $\delta \rho_1$ and $\delta \rho_2$:
\begin{equation}\label{adiabaticcondition}
    \frac{\delta \rho_1}{\rho_1 + p_1} = \frac{\delta \rho_2}{\rho_2+p_2}.
\end{equation}
Note that this result does not depend on the background metric considered.

Now we would like to consider the interacting case,
\begin{align}
    \nabla_M T_1^{MN} &= J^N, \quad \nabla_M N_1^M = j,\\
    \nabla_M T^{MN}_2 &= -J^N, \quad \nabla_M N_2^M = - j.
\end{align}
It should be clear from the discussion so far that, if the interaction is such that the continuity equations are unmodified and the individual particle number densities are conserved, then all the calculations above hold, and adiabatic perturbations satisfy \eqref{adiabaticcondition}. In other words, if $U_N J^N = 0 = j$, the condition for adiabatic perturbations in the interacting case is the same for the non-interacting case. 

Let's apply these discussions to cosmology. In standard cosmology, before matter-radiation equality, it is necessary to consider a fluid of interacting radiation and baryonic matter. The radiation interacts with the baryons via Thompson scattering. However, to first order, this interaction is such that there is no energy exchange between the components (only momentum transfer), and it also conserves particle number \cite{Baumann:2022mni}. This can be seen from the expression for $J^N$, $J^N \propto \sigma_T (U^N_\gamma + U^N_b)$, where $\sigma$ is the Thompson scattering cross-section, $U^N_\gamma$, and $U^N_b$ are the radiation and matter velocity fields \cite{Uzan:1998mc}. So, $J^N = 0$ at the background level, because both the radiation and matter fluids follow the Hubble flow to that order. Moreover, to first order in perturbation, $U^NJ_N$ still vanishes and only the perturbed Euler equations for the fluid get contributions from $J^N$, such that there is momentum exchange. Hence, we conclude that there is no dissipation or entropy production to leading order. If the initial perturbations were adiabatic, then we can neglect entropy perturbations entirely. Note that this is not the case if we go beyond the tight-coupling approximation.

Similarly, in the analysis of the previous section, if we assume that the bulk and brane fluid perturbations are adiabatic, we have
\begin{equation}
    \frac{\delta \rho_b}{\rho_b + p_b} = \frac{\delta \rho}{\rho+ p},
\end{equation}
on the brane. This corresponds to 
\begin{equation}\label{alphaadiabaticpert}
    \alpha = \frac{1+w_b}{1+w}
\end{equation} 
and the expression \eqref{soundspeed1}f or the effective sound speed can be written as
\begin{equation}
    c_{\text{eff}} = c_s \sqrt{\frac{1+ \frac{c_b^2}{c_s^2}R}{1+ \Bar{\gamma}R}} \;, 
\end{equation}
where we defined the \emph{bulk loading} $R$ as
\begin{equation}
     R = \frac{\rho_b (1+ w_b)}{\rho(1+w)}.
\end{equation}
Note that if $R>0$ (which is the case provided the bulk and brane fields satisfy the null energy condition), then $c_b^2 > \Bar{\gamma}c_s^2$ implies $c_{\text{eff}}>1$. 

\section{Adding expansion}
\label{sec4}

The goal of this section is to estimate how the cosmological expansion would change equation \eqref{pertonbrane}. Suppose we want to consider the dynamics of the two fluids in a spacetime with metric
\begin{equation}
    ds^2 = - dt^2 + a^2(t) dx^i dx_i + b^2(t)dy^2,
\end{equation}
in the brane comoving frame. We will not impose Einstein's equations, i.e., we are still assuming test fluids in a fixed curved background. A fully-fledged study of the cosmological perturbations in our setting is beyond the scope of this paper. Instead, we are interested in the sound horizon of the moving-brane fluid when it is coupled with the bulk fluid, and for that, we can neglect the backreaction in the geometry.

Computing the Christoffel symbols
\begin{equation}
    \Gamma_{ij}^0 = a^2 H \delta_{ij}, \Gamma_{0i}^j = H \delta^i_j, \quad \Gamma_{yy}^0 = b^2 H_b, \quad \Gamma_{04}^4 = H_b, 
\end{equation}
where $H = \partial_0 \ln a$ and $H_b = \partial_0 \ln b$, and using them into \eqref{branefluid} gives, for $U^M = (1,0,0,0,0)$,
\begin{subequations}
    \begin{align}
        \partial_0 \rho + (\rho+p)\left(3H + H_b\right) &= -J_0,\\
        a^{-2}\eta^{ij}\partial_j p &= J^i,\\
        (b^{-2}-1)\partial_4 p + \partial_4 p_4 &= J^4.
    \end{align}
\end{subequations}
For the bulk, with $U^M_b = \gamma(1, u^i_b, -v)$, equations \eqref{bulkfluid} give
\begin{subequations}
    \begin{align}
        \partial_0 \rho_b + u^i_b \partial_i \rho_b - v \partial_4 \rho_b + (\rho_b +p_b)(3H + H_b)+\frac{1}{\gamma}(\rho_b + p_b)\left[\partial_0 \gamma + \partial_i (\gamma u^i_b)-\right.\nonumber \\
        -\left.\partial_4 (\gamma v)\right] = J_0 + J_i u^i_b - J_4 v, \\
        a^{-2}\eta^{ij}\partial_j p_b + (\rho_b+p_b)\left[\gamma \partial_0(\gamma u^i_b) +\gamma u^j_b\partial_j(\gamma u^i_b) - \gamma v \partial_4 (\gamma u^i_b) +2H \delta^i_j \gamma^2 u^j_b \right] + \nonumber\\
        +\gamma u^i_b \left(\gamma \partial_0p_b + \gamma u^j_b \partial p_b - \gamma v \partial_4p_b\right) = -J^i - (J_0 +u^i_bJ_i - vJ_4)\gamma^2 u^i_b, \\
        b^{-2}\partial_4 p_b+ (\rho_b +p_b) \left[-\gamma \partial_0(\gamma v)- \gamma u^j_b \partial_j(\gamma v) + \gamma v \partial_4 (\gamma v) - 2H_b \gamma^2 v \right] -\nonumber \\
        -\gamma v \left(\gamma \partial_0 p_b + \gamma u^j_b \partial_j p_b - \gamma v \partial_4 p_b\right) = -J^4 + (J_0 + u^i_b J_i - vJ_4)\gamma^2 v.
    \end{align}
\end{subequations}

Similar as in the flat case, we can find the equations for the brane and bulk perturbations after setting $U^M \approx (1, u^i, 0)$ and treating $u_b^i$ in $U^M_b$ as linear order in perturbations. As in flat space, we have $\delta v= 0$ and $J^0 = 0 = J^4$. For the brane variables, evaluating the equations for perturbations in a background with $\Bar{p}_4 = 0$, $\partial_j \Bar{\rho} = 0 = \partial_4 \Bar{\rho}$, and $\Bar{p} = w\Bar{\rho}$, we get
\begin{subequations}
    \begin{align}
        \partial_0 \delta + (1+ w) \partial_i u^i +(c_s^2-w)(3H+ H_b)\delta &= 0,\\
        c_s^2 a^{-2}\eta^{ij}\partial_j \delta+(1+w)(\partial_0 u^i +2 H u^i) + \left[w H_b - w(1+w)(3H+H_b)\right] u^i &= \frac{\Theta}{\Bar{\rho}}(u^i - \Bar{\gamma}u^i_b),\\
        c_s^2(b^{-2}-1)\partial_4 \delta &= 0,      
    \end{align}
\end{subequations}
where we used the background equations and $\bar{p} = w\Bar{\rho}$. For the bulk, using the background solution with $\partial_i \Bar{p}_b = 0 = \partial_i \rho_b$, and $U^i_b = 0$, we get
\begin{subequations}
    \begin{align}
        (\partial_0 - \Bar{v}\partial_4)\delta_b + (c_b^2 - w_b)(3H + H_b)\delta_b + (1+w_b) \partial_i u^i_b  &= 0,\\
        c^2_b a^{-2} \eta^{ij}\partial_j \delta_b + \Bar{\gamma}^2(1+w_b)\left[(\partial_0-\Bar{v}\partial_4)u^i_b +2 H u^i_b - w_b(3H+H_b)u^i_b\right] &= -\frac{\Theta}{\Bar{\rho}_b}(u^i - \Bar{\gamma}u^i_b),\\
        c_b^2 b^{-2}\partial_4 \delta_b -\Bar{\gamma}^2 \Bar{v}c_b^2(\partial_0 - \Bar{v}\partial_4)\delta_b + 2H_b \frac{\Bar{\gamma}^2\Bar{v}}{w_b}\left(c_b^2 -w_b\right)\delta_b &=0.    
    \end{align}
\end{subequations}
Note that $c_b^2=w_b$ and $c_s^2= w$, but we left some terms proportional to $(c_b^2-w_b)$ and $(c_s^2- w)$ explicit to depict how adiabaticity affects the equations. 

In the tight-coupling approximation, the bulk Euler equation gives
\begin{equation}
     -\frac{\Theta}{\Bar{\rho}_b}(u^i - \Bar{\gamma}u^i_b) \approx c^2_b a^{-2} \eta^{ij}\partial_j \delta_b + \Bar{\gamma}^2(1+w_b)\left[(\partial_0-\Bar{v}\partial_4)u^i +2 H u^i - w_b(3H+H_b)u^i\right], 
\end{equation}
and plugging this into brane's Euler equation, we find
\begin{align}
    c_s^2\left(1+ \alpha \frac{\Bar{\rho}_b c_b^2}{\Bar{\rho}c_s^2}\right)a^{-2}\eta^{ij}\partial_j \delta  + \left[1+ \Bar{\gamma}^2\frac{\Bar{\rho}_b(1+w_b)}{\Bar{\rho}(1+w)}\right](1+w)\partial_0 u^i &+\nonumber\\
    \left\{wH_b + 2H (1+w) \left[1+ \Bar{\gamma}^2\frac{\Bar{\rho}_b(1+w_b)}{\Bar{\rho}(1+w)}\right] - \left[1+\Bar{\gamma}^2 \frac{\Bar{\rho}_b(1+w_b)w_b}{\Bar{\rho}(1+w)w}\right]w(1+w)(3H + H_b) \right\}u^i &= 0.
\end{align}
Taking $\partial_i$ of the equation above and combining it with the time derivative of the brane's continuity equation gives
\begin{align}\label{branedeltaeq}
    \partial_0^2 \delta - c^2_s\frac{1+ \alpha \frac{\Bar{\rho}_b c_b^2}{\Bar{\rho}c_s^2}}{1+ \Bar{\gamma}^2\frac{\Bar{\rho}_b(1+w_b)}{\Bar{\rho}(1+w)}}a^{-2}\nabla^2\delta &+\nonumber\\
    +\left\{\frac{w}{1+w}\frac{H_b}{1+ \Bar{\gamma}^2\frac{\Bar{\rho}_b(1+w_b)}{\Bar{\rho}(1+w)}} +2H -\left[1+ \frac{\Bar{\gamma}^2\frac{\Bar{\rho}_b(1+w_b)w_b}{\Bar{\rho}(1+w)w}}{1+\Bar{\gamma}^2\frac{\Bar{\rho}_b(1+w_b)}{\Bar{\rho}(1+w)}}\right]w(3H+H_b)\right\}\partial_0 \delta &= 0.
\end{align}
We see that, as expected, the time-dependence of the scale factors introduces friction terms in the equation for $\delta$, without modifications to the effective sound speed. After selecting a background evolution for $H$ and $H_b$, one can solve \eqref{branedeltaeq} using a WKB approximation. Generically, the result will be oscillations with amplitude modulated by a slow-varying function of the time. Hence, the main effect of the assumptions on the brane-bulk coupling is the effective sound speed and its associated sound horizon. 

\section{Discussion and Conclusion}
\label{conclusion}

In this letter, we studied the first implications of the coupling between brane and bulk fluids when the brane moves in a compact direction. Due to the periodicity of the extra spatial direction, certain bulk propagating signals will undergo a time advancement that leads to a more prompt information transmission when signals probe the bulk rather than remaining parallel to the brane. This gives rise to an effective superluminal propagation with respect to the events on the brane.

We considered a system of bulk and brane-coupled perfect fluids for effective superluminal propagation to manifest in the brane. Similar to the Thomson interaction between photons and baryons, we considered an energy-momentum transfer flux proportional to the difference between the brane and bulk velocity perturbations. In the tight-coupling approximation, we manipulated the continuity equation and the Euler equations of the perturbations to find the effective sound speed for brane adiabatic perturbations.

Since our calculations are valid for test fluids in fixed flat and expanding backgrounds and we have not considered metric perturbations in the equations, it is worth understanding in what regime they apply for brane-world cosmology models. Metric fluctuations would appear as source terms in equation \eqref{branedeltaeq}, and the system would close after considering the perturbed Einstein's equations. Similar to the study of the CMB photon-baryon fluid oscillations, the high-frequency modes of the brane fluid correspond to perturbations in scales small compared to the Hubble radius. For those modes, the slowly varying, long-wavelength metric fluctuations sourcing the equation for $\delta$ can be neglected, and equation \eqref{branedeltaeq} is a good approximation for the physics of the oscillations. Hence, the resulting sound horizon before the brane-bulk decoupling
\begin{equation}
    r_s^{\text{eff}} = \int dz \frac{c_{\text{eff}}(z)}{H(z)} = \int dz \frac{c_s}{H} \sqrt{\frac{1+ \frac{c_b^2}{c_s^2}R}{1+ \Bar{\gamma}R}}
\end{equation}
is robust against metric fluctuations. Whether or not this sound horizon can be larger than in the $\Lambda$CDM model depends on the factor in the square root. If the brane and bulk fluids have radiation and dust-like equation of state, respectively, then $c_{\text{eff}}<1$ because $\Bar{\gamma}>1$. This is also the case if both fluids have a radiation equation of state. So, for these situations, the tight-coupling approximation does not allow for a ``super-acoustic'' propagation of perturbations. This implies a solution to the horizon problem along the same lines as in varying speed of light\footnote{See \cite{Kiritsis:1999tx, Alexander:1999cb} for proposals to get a varying speed of light from string theory.} (VSL) \cite{Bassett:2000wj, Magueijo:2003gj, Magueijo:2008pm}, and tachyacoustic cosmologies \cite{Bessada:2009ns} seem only possible after going beyond the tight-coupling approximation or for a bulk equation of state satisfying $w_b> \Bar{\gamma} w$. Remarkably, this can be satisfied without violating the null energy condition for the brane and bulk fluids. Deep in the radiation domination, for instance, we would need $w_b> \bar{\gamma}/3>1/3$, and the bulk fluid do not even need to violate the strong energy condition.

There are some ways our analysis can be generalized. Brane-world solutions of five-\linebreak dimensional Einstein gravity have been discussed previously \cite{Lukas:1998yy, Lukas:1998tt, Binetruy:1999ut, Khoury:2001wf,Khoury:2001bz} (see also \cite{vandeBruck:2000ju} and references therein), although in the interval $S^1/Z_2$ or with a static brane, with no possibility of effective superluminal brane propagation. We leave a complete study of how these solutions are modified when the compact direction is periodic for future work.  Moreover, it would be interesting to investigate more general energy-momentum transfer fluxes $J^N$, including dissipation and entropy perturbations. On a more speculative note, one could investigate whether the sound horizon modification above can alleviate the Hubble tension.

A string theory embedding of the setup might require generalizing the effective superluminal propagation to brane motion in more general compact spaces. A potential issue for a string description is the coupling of branes to $p$-forms, which induces a charge cancellation condition in the compact manifold \cite{Polchinski:1998rr}. Solving this Gauss constraint with anti-branes might spoil the return of bulk-propagating signals to the brane since the signal might end in the anti-brane instead. However, one can use bulk fields, like fluxes, to cancel the brane charge, and thus, the $p$-form coupling is not an obstruction for a single (or stack) brane motion. Understanding how the moving brane backreacts in the bulk geometry is more challenging. 

\section*{Acknowledgments}

We thank Robert Brandenberger for many discussions and comments, and Stephon Alexander and Keshav Dasgupta for comments on the first version of the paper. H.B. also would like to thank Niayesh Afshordi and Ghazal Geshnizjani for discussions. H.B. was supported by the Fonds de recherche du Qu\'ebec (PBEEE/303549). Research at McGill is partially supported by funds from NSERC and the Canada Research Chair program.





\bibliographystyle{bibstyle} 
\bibliography{references.bib}






\end{document}